\documentclass[11pt,a4paper]{article}

\usepackage[pdftex]{color}
\usepackage{amssymb}
\usepackage{amsthm}
\usepackage{amsmath}
\usepackage{latexsym}
\usepackage{epsfig}
\usepackage{amscd}
\usepackage{graphicx}
\usepackage[pdftex, colorlinks=true, citecolor=green]{hyperref}
\usepackage{lscape}

\begin{document}

\bibliographystyle{plain}

\title{A multi-etiology model of systemic degeneration in schizophrenia}
\author{Anca R\v{a}dulescu, Applied Mathematics, University of Colorado at Boulder\\ UCB 526, Boulder, CO 80309-0526, \emph{radulesc@colorado.edu}}
\maketitle

\begin{align*}
       &\text{\hspace{6cm} {\bf Motto:} \small{``\emph{I did have strange ideas during certain periods of time}.''}}\\ 
       &\text{\hspace{14.5cm} \small{John Nash}}
\end{align*}	

\vspace{10mm}

{\small{We discuss the possibility of multiple underlying etiologies of the condition currently labeled as schizophrenia. We support this hypothesis with a theoretical model of the prefrontal-limbic system. We show how the dynamical behavior of this model depends on an entire set of physiological parameters, representing synaptic strengths, vulnerability to stress-induced cortisol, dopamine regulation and rates of autoantibody production. Malfunction of different such parameters produces similar outward dysregulation of the system, which may readily lead to diagnosis difficulties in a clinician's office. We further place this paradigm within the contexts of pathophysiology and of antipsychotic pharmacology. We finally propose brain profiling as the future quantitative diagnostic toolbox that agrees with a multiple etiologies hypothesis of schizophrenia.}}

\section{Introduction}

\subsection{Schizophrenia. Disease or syndrome?}

As defined today, schizophrenia is a serious chronic disorder, affecting about 1.1$\%$ of the population (over 65 million people worldwide, according to NIMH statistics). It is a source of unrelenting personal and social drama, and underlies debilitating problems from unemployment and long-time hospitalization to suicide. 

As much as it would be beneficial to understand this disease and eliminate the torment it brings to human life, its cause is still unknown by research and its current treatment brings no reliable cure. The definition of schizophrenia itself is elusive, having been somewhat fabricated by science~\cite{DSM} --- based on statistical behavior, rather than etiology --- so as to somehow encompass all the complexities of a syndrome otherwise intractable. Progress has been made in understanding the effects of particular antipsychotics on some of the modules involved in the disease process, such as the dopaminergic or serotonergic brain systems, but the drugs that are being used may still treat the effects of the disease rather than its mysterious cause~\cite{Kap}.

Indeed, as of today, the etiology of schizophrenia is still not understood. It clearly affects structurally and functionally various cortical and subcortical regions involved in cognitive, emotional and motivational aspects of behavior~\cite{Law}~\cite{Anan}~\cite{Staa}~\cite{Andr}. However, it seems that the days when schizophrenia was labeled as ``brain sickness'' are over. Its disease process has been shown to have much more extensive physiological effects, including endocrine~\cite{Rits} and autonomic~\cite{MPY} factors, and has been correlated to a plethora of somatic abnormalities, affecting aspects as thorough as autoimmunity mechanisms, and as simple as ``eating your vitamins.''

In fact, ``schizophrenia's'' heterogeneous set of symptoms (hallucinations, delusional beliefs, thought disorder, emotional flattening, social withdrawal) is diverse enough to possibly represent a whole collection of diseases in their own right~\cite{BT}. In addition, most of the illness' signs and symptoms are not unique to schizophrenia and overlap with the symptoms of other mental conditions, such as major depression, bipolar mania, or post-traumatic stress disorder. In this context, newer research is gradually moving in the direction of altering the existing unreliable terminology, even towards entirely replacing it by a physiology-based system, optimizing predictive power.

Altogether, the diversity in symptomatology and the intractability of the disease's etiology could in fact have a \emph{bona fide} explanation: there are \emph{multiple etiologies} of schizophrenia~\cite{Crow}. It seems sensible to hypothesize that the cognitive and behavioral symptoms of schizophrenia are perhaps a \emph{generic} outward psychopathological manifestation -- much like ``heart condition''  in the somatic context -- which could be produced and maintained by any of an entire collection of causes and mechanisms. Each etiology, although producing similar symptoms, has however a different prognosis and should have a different diagnosis and most certainly a different treatment. This is why it is clinically important to make these distinctions, and to strive for a novel organization of the psychiatric terminology (see Section 4.3).

\subsection{Vulnerability and limbic dysregulation}

Significant research over the past decade has isolated the amygdala, the hippocampus and parts of the prefrontal cortex as the brain areas most relevant in the mechanisms of fear conditioning and emotion processing. 

It has been proposed that schizophrenic symptoms constitute an end-stage of a cyclic and neurodegenerative process, in which a hereditary predisposition (``vulnerability''~\cite{Nuech}) reduces the individual psychological threshold towards stimuli~\cite{Stamm}, to the point where even minor daily stresses will directly trigger psychotic experiences~\cite{Myin}. It has been further hypothesized and documented that the critical element in the physiological realization of this degeneration is dysregulation of limbic regions. Specifically, this manifests as an imbalance between the two components of a feed-back loop: (1) the top-down control provided by the cortical systems over subcortical areas involved in emotion and (2) the control of neuromodulatory systems that in turn affect the cortex. For example, one possible view involves the ability of the hippocampus and prefrontal cortex to contextualize and inhibit activation of the amygdala, the excitatory module~\cite{Sotres2004}, which in turn drives the dopaminergic system. Outputs of the dopamine system modulate processing in the hippocampus and PFC, thus closing the feed-back cycle (see Section 2). In this context, it is perhaps not surprising that schizophrenia has remained hard to understand, as a system governed by complex limbic interactions.

One line of current thinking is that this vulnerability to stress in schizophrenia is based on a pre-existing prefrontal-limbic deficit~\cite{Medoff}~\cite{Preston}~\cite{Tamm}. More precisely, an impairment in hippocampal/prefrontal function may lead to decreased inhibition of the amygdala and subsequent higher arousal levels, even under minor stress. Through the hypothalamo-pituitary axis, the fear reaction triggers autonomic and endocrine effects~\cite{LeDoux2003}, in particular increased cortisol levels ~\cite{Sap}. These produce brain neurotoxicity~\cite{Weinberger} and further hippocampus damage~\cite{Pavlides}, at the level of neurogenesis~\cite{Herbert}~\cite{Wong} and of hippocampal--PFC synaptic transmission~\cite{Cerqueira}. This is possibly one of the main degenerative cycles in schizophrenia, which the author has discussed in previous work~\cite{Radulescu}.

This cortisol neurotoxicity hypothesis is gradually gaining ground in current approaches to mental pathology~\cite{Walker}. However, other mechanisms of limbic vulnerability have been proposed by biological and clinical studies, whose bases do not necessarily reside within the physiology of the brain. The limbic dysregulation (and subsequent outward psychosis) could be the result of a multitude of causes, including genetic confounds, drug abuse, brain trauma, or even lack of proper minerals from food (such as Vitamin $B_{12}$ and folate~\cite{Ozcan}~\cite{Haidemenos}). These factors can all work towards weakening the limbic parameters to a level which places the system in a mode of enhanced vulnerability to environmental pressure. 

One of the better known such factors is autoimmunity (i.e. the inability of the body to suppress production of antibodies aimed against its own structures). In this paper, we extend our previous model of limbic vulnerability to include the effects of autoimmunity on the prefrontal-limbic system. After a short introduction on autoimmunity and psychiatric manifestations of autoimmune conditions, we organize the rest of the paper as follows. 

Section 2 describes the brain pathways and physiological mechanisms included in our paradigm, and constructs the mathematical model. Section 3 illustrates, using MATLAB simulations, the behavior of the model and its evolution in a few cases of interest. Section 4 relates this mathematical behavior with the brain pathophysiology, and places it within the context of identifying different etiologies for schizophrenia. Finally, Section 5 takes a glimpse into a possible future of psychiatric diagnoses and treatment, if based on such temporal architecture modeling.

\subsection{Autoimmunity --- a road to psychosis}

Basic and clinical experimental data has correlated autoimmune dysregulation to autward psychosis. To fix our ideas, we will consider a standard example of a systemic autoimmune condition: Systemic Lupus Erithematosus (SLE). It affects mainly young people by attacking indiscriminately --- with sudden, seemingly unpredictable flares --- multiple organ systems, among which the brain. Neuropsychiatric manifestations have a prevalence of up to $75-90\%$ in SLE. However, despite this strong correlation between autoimmune deficiencies and central neurological and psychiatric problems, studies have not consistently proved any causality between the two phenomena, in either direction. It is quite possible that the disease system contains bilateral interactions between the immunity and central nervous subunits~\cite{Bongioanni}, as well as other external modulations (e.g., from external stress~\cite{Gaughran} and the subsequent hypercortisolemia). 

Although not unexpected, this complexity complicates a possible global analysis that would lead to better understanding of the system. On the other hand, this understanding is clinically very important. SLE symptoms ---  including depression, anxiety, psychosis, mania, cognitive dysfunction --- have been more than once ``misdiagnosed" as schizophrenia in a psychiatrist's office~\cite{page}. Clearly, it is of crucial importance to avoid confusions, since in such a case a treatment trial should be planned to match the physiological autoimmune condition, not just to alleviate the outward psychiatric symptoms with antipsychotic medication~\cite{Cohen}. 

As we have already mentioned, in autoimmune conditions such as SLE, the body produces autoantibodies, i.e., antibodies to its own tissues. It is thought that most forms of autoimmunity start up for no obvious reason, apparently by chance~\cite{Edwards}, with the production of an autoantibody which, through a vicious cycle, stimulates its own production. One such example in SLE is the autoantibody that targets a complement system molecule called C1q, causing decreased apoptotic cell clearance, and thus facilitating more autoantibody production~\cite{Scolding}. This and other complexities suggest that lupus may be driven by a whole series of connected such antibody cycles, possibly themselves generated by low C1q levels. 

Among the variety of autoantibodies produced in SLE, some are particularly dangerous to the central nervous system (CNS). Several studies have focused on isolating antibodies that target specific brain structures, such as neuronal membrane antigens, ribosomal proteins or endothelial surfaces. For example: anti-cardiolipin (an antiphospholipid) produces prothrombotic effects which lead to ischemia and infarction and eventually to neuronal death~\cite{Gharavi}. Aditionally, anti-endothelial antibodies~\cite{Valesini} contribute to the general damage of vessel wall endothelium, thus lowering the blood-brain barrier and allowing easier penetration of other antibodies or neurotoxic factors from the blood into the CNS. Although most molecular mechanisms of CNS-directed autoimmunity are still under investigation, it is believed that, rather than irreversible cell death, some anti-neuronal antibodies~\cite{Bresnihan} may have more transient detrimental effects on neural function. These effects include signaling, myelination~\cite{Kipnis}~\cite{Jov}, synaptic plasticity, neurotransmitters~\cite{Moskowitz} and receptor dynamics.

The antineuronal antibodies most typical to lupus are anti-DNA antibodies, which act at the level of NMDA receptors~\cite{Omdal}. Abnormal synaptic remodeling due to dysregulated NMDA receptor trafficking provides a very plausible mechanism for neuropsychiatric disorders such as Alzheimer's disease, cocaine addiction and schizophrenia~\cite{Zhang}~\cite{Lau}. This is why in our model the effects of autoimmunity will be introduced at the synaptic level, as an impairment in transmission in certain neural pathways.
Since our study is concerned with the brain dynamics between prefrontal and limbic regions of interest, understanding if and how they are targeted by autoimmunity is very useful. In this sense, we use recent results on the possible selective production of antibodies against specific brain regions~\cite{Rothermundt}. Indeed, serum antibodies have been reported against a few regions, including the hippocampus~\cite{Ganguli}~\cite{Kelly}~\cite{Yang}, the amygdala~\cite{Henneberg} and frontal cortex~\cite{Henneberg}. 

Finally, it is very important to notice in our context that the classical SLE treatment plan is based on corticosteroid medication, due to its immunosuppressive effects~\cite{Denburg}. Severe diffuse CNS manifestations, such as acute confusional state, generalized seizures, mood disorders and psychosis, generally require immediate corticosteroid administration~\cite{Sanna}, and psychiatric symptoms tend to resolve within three weeks of treatment. On the other hand, hypercortesolemia (possibly through the neurotoxic effects on the prefrontal/hippocampal inhibitory module discussed earlier in the section) could itself cause a variety of psychiatric syndromes such as mania, depression, psychosis, anxiety, dellirium~\cite{Cohen}. It seems therefore crucial that a medication plan be finely tuned to correspond to the particular needs of each individual patient, since an overdose could lead to exactly the undesired symptoms that it aims to cure. This emphasizes yet again the importance of a sustainable and detailed enough diagnosis that would permit a correct treatment evaluation.

In the Discussion section, we will further interpret the potential  interference of corticosteroid medication with the central and endocrine mechanisms that are beleived to be responsible for psychosis in schizophrenia.

\section{The construction of the mathematical model}

\begin{figure}[h!]
\begin{center}
\includegraphics[scale=.3]{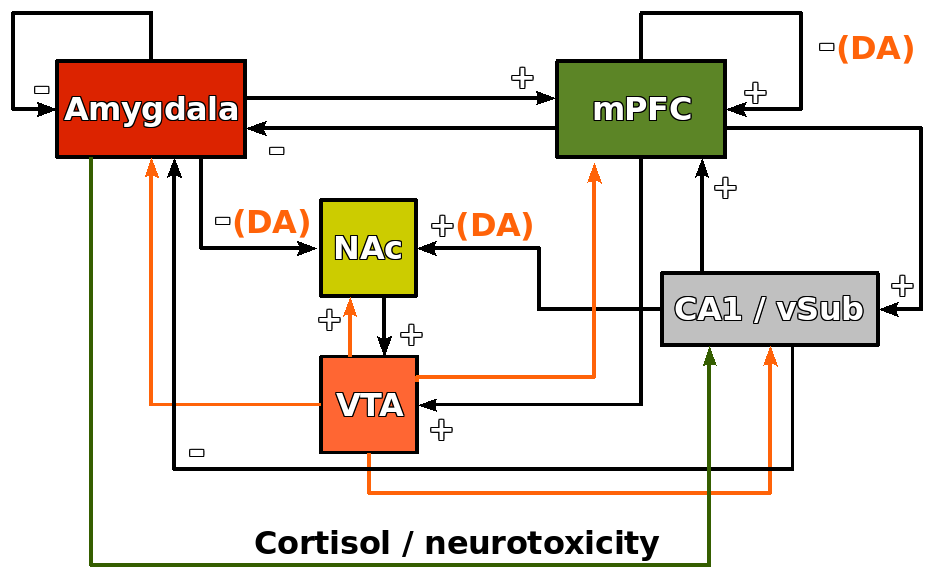}
\caption{Simplified schema of the prefrontal-limbic pathways incorporated in our model, as described in Section 2. mPFC = medial prefrontal cortex; NAc = nucleus accumbens; VTA = ventral tegmental area; CA1 = Cornu Ammonis 1; vSub = ventral subiculum; ``$+$'' stands for excitatory pathway, ``$-$'' for inhibitory pathway and ``DA'' for dopamine modulated pathway.}
\end{center}
\end{figure}

In our previous work~\cite{Radulescu}, we have related schizophrenia to a vulnerability of the inhibitory limbic/prefrontal module reinforced by neurotoxic effects of stress-induced hypercortisolemia. In this section we will describe a mathematical model which, while still based on this limbic vulnerability, will also reflect some of the more subtle physiological mechanisms discussed in the previous sections. If constructed correctly, this extended model will address in a biologically plausible way not only the underpinnings of fear conditioning and extinction, but also the impact of autoimmunity on limbic processing and the effects of corticosteroids and antipsychotic medication which are generally prescribed for SLE and schizophrenia. 

We start with a simpler case, not including the effects of autoimmunity. First, we define our terminology and notations, and we elaborate the simplified model. We then continue by explaining the underlying set of pathways and physiological rules on which we based our paradigm. 

We represent the time activations of the amygdala, the hippocampus and the prefrontal cortex as three distinct variables $a$, $p$ and $h$, while a fourth variable $\delta$ stands for the activation of the dopamine system, controlled via the nucleus accumbens (NAc) and the ventral tegmental area (VTA). The interactions between these variables are quantified by a collection of real positive parameters $I, M, \mu_1, \mu_2, k_1, k_2, k_3, \gamma_1, \gamma_2, a_1, a_2, \alpha$, $\beta$, $\xi_1$, $\xi_2$ and $\xi_3$ as follows:

\begin{align*}
\dot{a} &=-\mu_1 a - k_1 p-\gamma_1 h + I\\
\dot{p} &=k_2 a - \mu_2(1+\delta) p+\frac{\gamma_2}{a_1 C+1} h\\
\dot{h} &=k_3 (1+\delta^2) p - a_2 C\\
\dot{\delta} &=-\xi_1 (1+\delta)a+\xi_2(1+\delta)p+\xi_3(1+\delta)h
\end{align*}

Aside from a self-modulatory inhibition $-\mu_1 a$, the amygdala receives a constant environmental input $I$ via thalamic pathways, and provides excitatory outputs $k_2 a$ to the prefrontal cortex~\cite{Sotres2004}. Newer findings indicate that BLA receives a stress-responsive dopamine (DA) projection from the ventral tegmental area (VTA)~\cite{Lisman}, and itself dampens the DA response in the NAc~\cite{Stevenson}. Outputs from the amygdala, through the hypothalamo-pituitary axis, also provide inputs for the endocrine and autonomic nervous systems, controlling indirectly the cortisol production in response to stress~\cite{Sap}.

In our model, we express the amygdala-controlled cortisol as a function $\displaystyle{f(\alpha,a)=\alpha \frac{e^{a}}{e^{a}+1}}$, so that any activation $a$ of the amygdala produces an increase from zero in the production of cortisol. This increase cannot exceed a saturation level $\alpha$, independently of the level of stress. The stress-induced cortisol, together with corticosteroid-based medication $M$, determine the levels of blood cortisol $C=M+f(\alpha,a)$, whose known neurotoxic effects influence the structure and activity of the hippocampal/prefrontal inhibitory unit, as shown below.

The prefrontal cortex receives excitatory inputs $k_2 a$ from the amygdala and strong excitatory hippocampal projections $\gamma_2 h$ to the prefrontal prelimbic and infralimbic regions. The synaptic efficacy of this pathway is impaired by increased levels of blood cortisol; to account for this, we have adjusted the corresponding term to $\displaystyle{\frac{\gamma_2}{a_2 C+1} h}$. More recent research reveals the importance of dopaminergic modulation of the mPFC inhibition via GABA interneurons. Projections from the mPFC to the VTA~\cite{Brady}, together with a reciprocal dopaminergic pathway from the VTA to the mPFC, comprise the mesocortical circuit~\cite{Harte}~\cite{Westerink}. More precisely, the glutamate-containing pyramidal neurons which are the target of VTA DA terminals make synapses on non-pyramidal GABAergic inhibitory interneurons in the mPFC~\cite{Harte}~\cite{Seamans}. Interneuron inhibition in mPFC has been related to schizophrenia in recent studies~\cite{Volk}. We have expressed this dopamine modulated self-inhibition as the term $-\mu_2 (1+\delta) p$.

The hippocampus directly inhibits activation of the amygdala ($-\gamma_1 h$) and reinforces activity in the prefrontal cortex, supporting the process of memory formation and centralization ($\gamma_2 h$). In this process, the contribution of the DA system is again very important. The CA1/ventral subiculum (vSub) send dense glutametergic projections to the NAc, which are under a powerful DA neuromodulatory influence exerted by the VTA~\cite{Grace}. NAc directly projects to the VTA, and, in turn, the dopaminergic input from the VTA is knwon to enhance long term potentiation in CA1~\cite{Lisman}. In addition, the hippocampus receives modulations from both the amygdala (cortisol-neurotoxic inhibition $-\gamma C$) and the mPFC (excitatory modulation throught multineuronal etnorhinal and perirhinal pathways $k_3 (1+\delta^2) p$, where the factor $(1+\delta^2)$ signifies the dopamine-modulated LTP enhancement in CA1). 

The dopamine system receives dopamine-modulated  inputs from all three prefrontal-limbic areas, as shown by the fourth equation, in which the strengths of these modulations are proportional to the parameters $\xi_1$, $\xi_2$ and $\xi_3$.

As mentioned in the introduction, the large variety of autoantibodies and the complexity of their effects are not very well differentiated by research, and are therefore difficult to encompass in all detail. So, in order to conveniently extend our model to encompass autoimmunity, we focused on its effects at the level of amygdalar and hippocampal synaptic transmission, since these structures and areas have been particularly shown to be affected by autoimmunity in SLE. We consider the level of anti-neuronal autoantibodies to be modeled by an exponential increase in time: $B(t) = e^{\beta t/C}$. Here $\beta$ is the exponential growth rate that reflects the self-enforcing vicious cycle of autoantibody production, and $C$ is the blood level of cortisol, which slows down (without completely stopping) this production. With these additions, the system becomes:

\begin{align*}
\dot{a} &=-\mu_1 a - k_1 p-\frac{\gamma_1}{e^{\beta t/C}} h + I\\
\dot{p} &=\frac{k_2}{e^{\beta t/C}} a - \mu_2 p+\frac{\gamma_2}{e^{\beta t/C}(a_1 C+1)} h\\
\dot{h} &=k_3 (1+\delta^2) p - a_2 C\\
\dot{\delta} &=-\xi_1 (1+\delta)a+\xi_2(1+\delta)p+\xi_3(1+\delta)h
\end{align*}

Notice that, through the parameters $\mu_1$, $\mu_2$, $k_1$, $k_2$, $k_3$, $\gamma_1$, $\gamma_2$, $\xi_1$, $\xi_2$, $\xi_3>0$, the brain region interactions, represented by the linear terms of the system, reflect the excitatory/inhibitory character described in Section 2.1. The finer and slower modulations due to the effects of cortisol, dopamine or CNS antibodies are represented here as nonlinearities, consistent with our previous work~\cite{Radulescu}.

\section{Analytical results and simulations}

Generally, the mathematical analysis of such a system could be cumbersome, even intractible. Considering the large number of parameters that can be varied independently, we expect the behavior of the system to change with these parameters' variations and exhibit complex bifurcations at particular critical values. 


In future work, we plan to focus around such detailed analyses, and on numerical computation of critical bifurcation parameters when theoretical approaches are not possible. However, the purpose of this paper is not to build a complete mathematical analysis of the model we have constructed, especially since the model itself is not uniquely defined physiologically (most pathways, modulations and parameter values are still under investigation). We will instead adopt here a less precise, but rather more descriptive approach, aimed at illustrating basic ideas, such as the possible multitude of etiologies that underlie schizophrenic outward symtoms.   

In this section, we fix a set of plausible initial states, and we simulate the time evolution of these states under particular combinations of parameter values. Since we are mainly interested in the levels of prefrontal-limbic activations over time, and since we had to pick a 3-dimensional subspace for our graphic illustrations -- we plotted all trajectories in the 3-dimensional phase-space $(a,p,h)$. We observe in particular how these trajectories distort as the parameters change in a controlled way. Although we do not investigate analytically the phenomena that appear under deformations of the system, the simulations suggest the presence of multiple Hopf and Bautin bifurcations as parameters traverse certain tresholds.

\subsection{Model behavior without autoimmunity}

In our previous work, we have analyzed how the behavior of a two-dimensional prefrontal-limbic system changes under particular changes of the system's parameters. As a start, we wanted to verify that this extended paradigm was indeed consistent with our prior modeling results. Therefore, we first focused on investigating the behavior of the trajectories under perturbations of the amygdala self-inhibition strength $\mu_1$ and of the prefrontal/hippocampal vulnerability to cortisol neurotoxiticy, represented here by $a_1$ and $a_2$.

\begin{figure}[h!]
\begin{center}
\includegraphics[scale=.45]{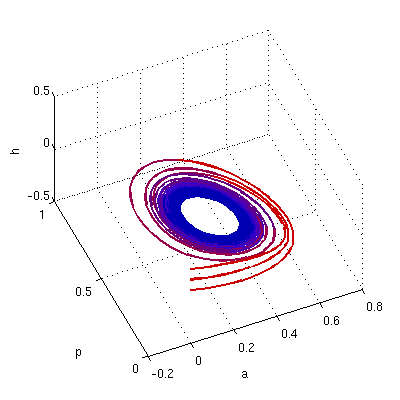}
\includegraphics[scale=.45]{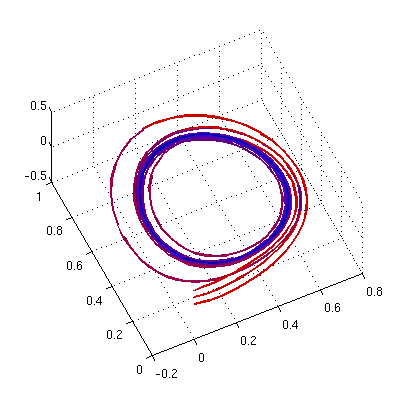}
\includegraphics[scale=.45]{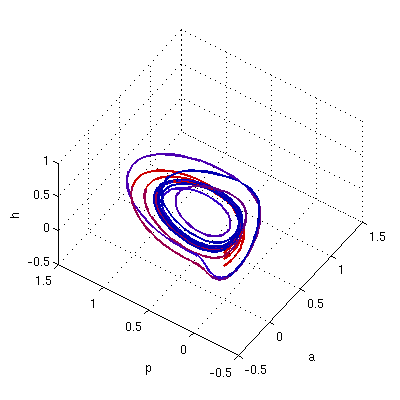}
\caption{\emph{The three panels show the evolution of the three fixed trajectories in the 3-dimensional phase-subspace $(a,p,h)$. We have fixed: $\mu_2=2$, $k_1=2$, $k_2=2$, $k_3=1$, $\alpha_1=1$, $a_1=a_2=1$, $\xi=2$, $I=1$, $\alpha=0$ and $M=0$ (i.e. no autoimmunity). {\bf A.} $\mu_1=1.4$; the trajectories are attracted to a point. {\bf B.} $\mu_1=1$; the attractor is switching from a point to a cycle. {\bf C.} $\mu_1=0.21$; the cycle has disappeared and the trajectories are no longer attracted.}}
\end{center}
\end{figure}

We gradually decreased the amygdala self-inhibition $\mu_1$, while keeping the other parameters fixed. As expected, large enough values of $\mu_1$ forced the time-evolutions to converge to a steady state. As we lowered $\mu_1$, we crossed a critical value where the trajectories seemed to stabilize to a locally attracting cycle. For values lower than this critical value, the attractor was lost, and the trajectories became more chaotic (see Figure 2). 

A similar bifurcation appeared to manifest when varying the parameters $a_1$ and $a_2$. Indeed, while for small values of these two parameters the system converged to a stable state, increases in either one of the two values led to the shift of the attractor state to an attracting cycle, and then to total loss of the local attractor, which caused the trajectories to be repelled away from the steady state.

\begin{figure}[h!]
\begin{center}
\includegraphics[scale=.45]{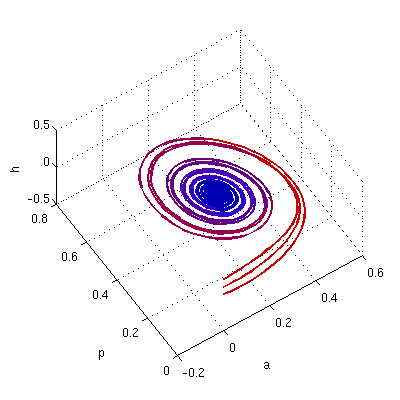}
\includegraphics[scale=.45]{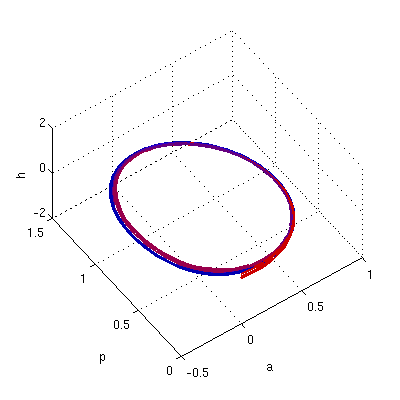}
\includegraphics[scale=.45]{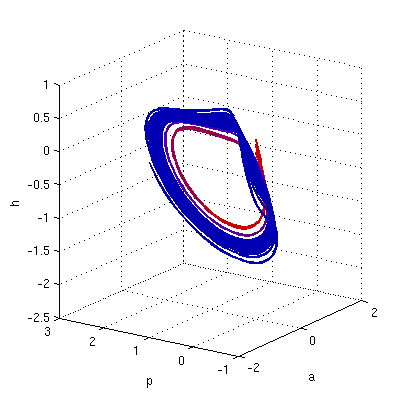}
\caption{\emph{In all three panels, we fixed: $\mu_1=2$, $\mu_2=2$, $k_1=2$, $k_2=2$, $k_3=1$, $\alpha_1=1$, $\xi=2$, $I=1$, $\alpha=0$ and $M=0$ (i.e., no autoimmunity). {\bf A.} $a_1=a_2=1$. The trajectories are attracted to a point. {\bf B.} $a_1=5$ and $a_2=2$. The trajectories are attracted to a cycle. {\bf C.} $a_1=5$, $a_2=2.64$. The trajectories are repelled.}}
\end{center}
\end{figure}

Before bringing autoimmunity into the picture, one last interesting behavior we wanted to understand was the effect of the dopamine levels / dopamine receptivity in the limbic and prefrontal modules of this model. As we increased the prefrontal-limbic modulation parameters of the dopamine system from $\xi_1=\xi_2=\xi_3=\xi=2$ to $\xi_1=\xi_2=\xi_3=\xi=2$, the system became more unstable, as shown in Figure 4. The Discussion section offers further interpretations of this loss of stability and of its significance in the context of antipsychotic medication.

\begin{figure}[h!]
\begin{center}
\includegraphics[scale=.45]{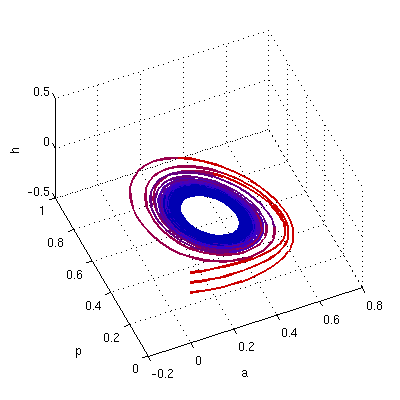}
\includegraphics[scale=.45]{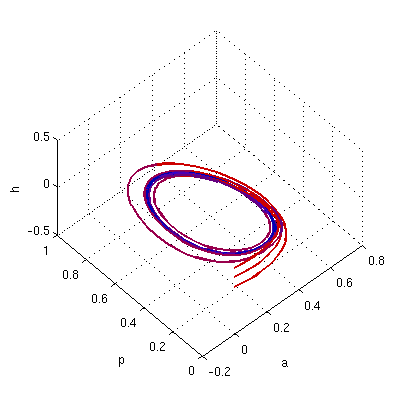}
\includegraphics[scale=.45]{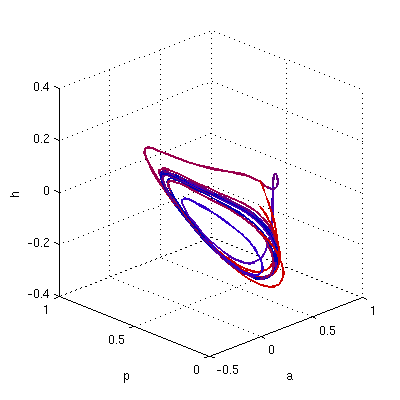}
\caption{\emph{In all three panels, we fixed: $\mu_1=1$, $\mu_2=2$, $k_1=2$, $k_2=2$, $k_3=1$, $\alpha_1=1$, $a_1=a_2=1$, $I=1$, $\alpha=0$ and $M=0$ (i.e. no autoimmunity). {\bf A.} $\xi=2$; the trajectories are attracted to a point. {\bf B.} $\xi=3$; the trajectories are attracted to a cycle. {\bf C.} $\xi=3.5$; the trajectories are repelled.}}
\end{center}
\end{figure}

\subsection{Model behavior including autoimmunity}

The synaptic impairment due to autoimmunity, as introduced in this model, depends directly on two quantities: the exponential growth rate of antibody production $\beta$ and the blood levels of cortisol $C$, which acts as an immunosuppressor. Figures 5 and 6 illustrate how corticosteroid medication works to reduce the psychiatric effects of autoimmunity.

For all panels of Figure 5, the growth rate was fixed to $\alpha=0.01$. Figure 5a shows that, when no medication was administered ($M=0$), the trajectories increased indefinitely along an invariant direction, so there could be no stable state. When medication was increased to $M=3$, the system locally stabilized, so the effects of autoimmunity were at least temporarily counteracted (Figure 5b). When medication was increased over a certain threshold, the neurotoxic effect of cortisol prevailed over its immunosuppressive effect, and the system became again unstable. This is shown in Figure 5c, where the medication has been increased to $M=4.165$, which was too high for the respective antibody growth.

\begin{figure}[h!]
\begin{center}
\includegraphics[scale=.45]{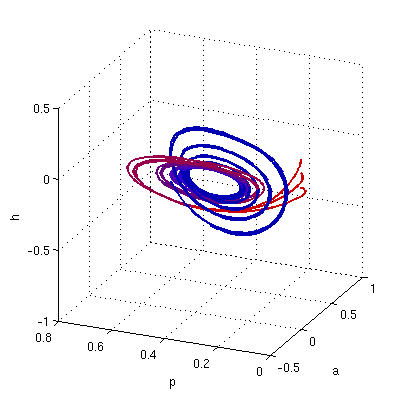}
\includegraphics[scale=.45]{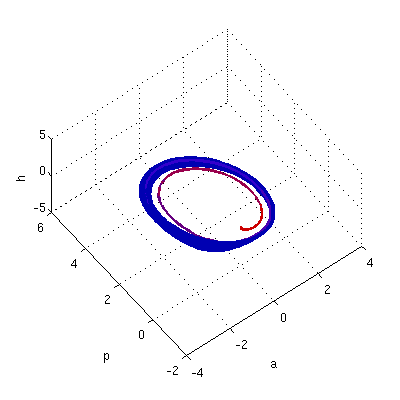}
\includegraphics[scale=.45]{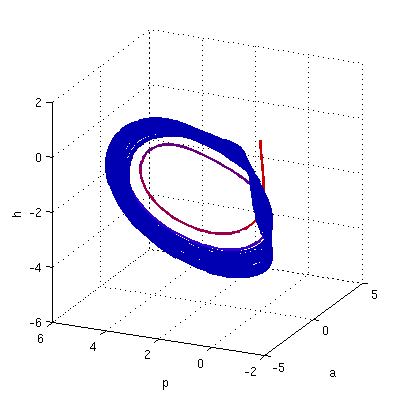}
\caption{\emph{In all three panels, we fixed: $\mu_1=2$, $\mu_2=2$, $k_1=2$, $k_2=2$, $k_3=1$, $\alpha_1=1$, $a_1=a_2=1$, $\xi=2$, $I=1$, $\alpha=0.01$. {\bf A.} $M=0$; the trajectories are attracted to a point. {\bf B.} $M=3$; the trajectories are attracted to a cycle. {\bf C.} $M=4.165$; the trajectories are repelled.}}
\end{center}
\end{figure}

Figures 6a and 6b illustrate the same phenomena for a slightly higher growth rate $\alpha=0.05$. The medication $M=3$, which was sufficient to stabilize a system with $\alpha=0.01$, is no longer sufficient, and the dose has to be increased up to $M=6$ to regain stability. A slightly larger dose, however,  would be again no longer useful (not shown).

\begin{figure}[h!]
\begin{center}
\includegraphics[scale=.45]{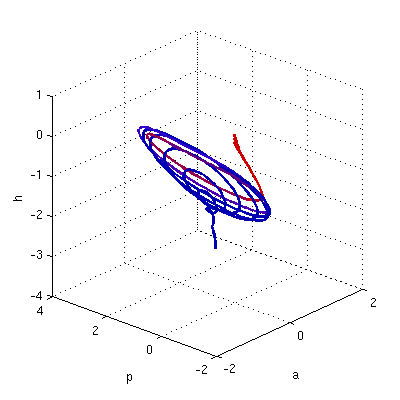}
\includegraphics[scale=.45]{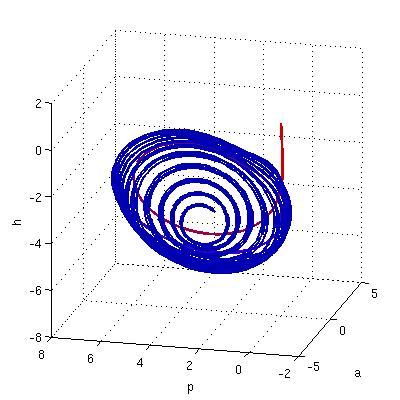}
\includegraphics[scale=.45]{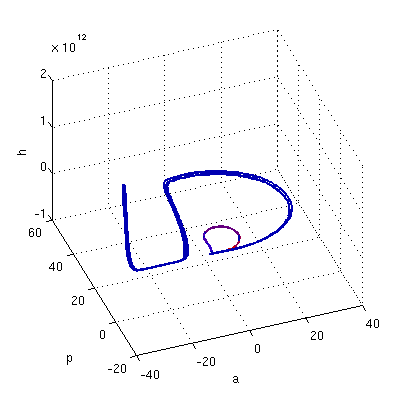}
\caption{\emph{In all three panels, we fixed: $\mu_1=2$, $\mu_2=2$, $k_1=2$, $k_2=2$, $k_3=1$, $\alpha_1=1$, $a_1=a_2=1$, $\xi=2$, $I=1$. {\bf A.} $\alpha=0.05$, $M=3$; the trajectories are repelled. {\bf B.} $\alpha=0.05$, $M=6$. The attractor is recovered. {\bf C.} $\alpha=0.1$, $M=15$. The trajectories can't be forced to converge even with very large values of $M$.}}
\end{center}
\end{figure}

It seems that for very large growth rates, this critical range in which medication stabilizes the system disappears. This is suggested in Figure 6c: for $\alpha=0.1$, the medication has been increased up to $M=15$, with no beneficial effect on the system's behavior at any intermediate stage.

We finally noted that, unlike in the original model (Figure 3), in the extended model decreasing $\xi$ does not force the trajectories to converge (not shown). In other words: while a lower drive of the dopamine systems had a stabilizing effect in the simplified model, in the extended model a similar decrement in dopamine modulation could not compensate for the effects of dysregulated autoimmunity. On the other hand, a slight increase in cortisol levels seems to do that successfully, as long as the dysregulation is not excessively pronounced. In the Discussion section, we offer a more detailed explanation of this phenomenon within the context of antipsychotic and immunosuppresant medication.

\section{Discussion}

Our model shows how a brain network involved in emotional processing, fear conditioning and extinction can switch from normal to dysregulated behavior when certain functional parameters are changed. More importantly, our paradigm shows how seemingly similar signs of dysregulation could correspond to malfunction of totally different parameters, i.e., they could be manifestations of totally different types of physiological impairment. This goes hand in hand with our hypothesis of multiple etiologies of psychosis, as described in the Introduction section. We would therefore like to interpret here more thoroughly our results that support this idea.

In our previous two-dimensional approach~\cite{Radulescu}, we have already discussed some of the effects of perturbing the amygdalar self-inhibition $\mu_1$ and the cortisol vulnerability coefficients $a_1$ and $a_2$. While we associated the linear coefficient with the individual's trait anxiety, the vulnerability parameter reflected the degree of the nonlinearity, which we believe to be the key determinator of the disease process. Our new computations confirm and extend these prior results. The loss of stability with the decrease of amygdalar self-inhibition reaffirms the importance of this inhibition in maintaining a well-regulated limbic-prefrontal system. Lack of proper such inhibition may lead to a less efficient return to baseline after a stressor. On the other hand, a dysregulation in the inhibitory feedback loop produced by increased vulnerability to stress (larger $a_1$ or $a_2$) may cause time evolutions to never converge, thus leading to ranges of brain activations compatible with psychotic behavior~\cite{Radulescu}. 

As an extension of this paradigm, we introduced here explicitly the effects of the dopamine modules, which are known to be very important in the genesis of psychosis. We observed that an increase in the dopaminergic modulation could produce a dysregulation in the system's evolution which is almost indistinguishable from an increase in vulnerability to cortisol. This effect corresponds to the well-known fact (which lies at the very basis of the pharmacology of antipsychotics~\cite{Awouters}) that dopamine antagonists alleviate (without definitively curing) psychotic symptoms. In the same direction, we also noted that a dysregulation in our system caused by increased prefrontal/hippocampal cortisol vulnerability can be compensated by decreasing the dopamine responsiveness, i.e., by administration of dopamine antagonist medication. This is also in full accordance with the theory of existing medical treatment for schizophrenia. 

The rest of the discussion will focus on the impact of autoimmunity. The game of autoimmune regulation is governed by the exponential growth rate of the autoantibody production. This rate can be effectively decreased by boosting the blood cortisol levels, e.g. by administration of corticosteroids. This slows down the exponential growth and implicitly lowers the detrimental effects of the antibodies on the brain. The corticosteroid treatment of defective autoimmunity works only to some degree, since an excess of cortisol can push the system into the opposite range of instability, due to hypercortisolemia. Indeed, recent studies have revealed that psychosis in SLE can appear as a secondary effect of medication, and that this psychosis is clinically indistinguishable from primary SLE psychosis~\cite{Cohen}.

In general, however, corticosteroid medication works well if administered correctly in SLE~\cite{Shanahan}. A very important message of our model is that, in the case of an actual autoimmune disorder, antipsychotics can't be regarded as a substitute or alternative to immunosuppressants. This emphasizes once more the importance of teasing apart the different mechanisms underlying psychosis, and of assigning medication customized to fit the particular corresponding malfunction. In our case, misdiagnosing SLE as schizophrenia, which probably happens more often than we know~\cite{page}, would lead to administering classical antipsychotic medication to a person suffering of an autoimmune condition. This is very ineffective, and detrimental to health, given the side effects of all antipsychotics. (Note on linguistics: To be able to express some distinctions, we confine ourselves here to refer to ``schizophrenia'' in its classical sense of stress/vulnerability condition.) Fortunately, the reciprocal mistake of misdiagnosing schizophrenia as SLE does not happen that often, since SLE usually comes with more identifiable, nonpsychiatric signs and symptoms. The consequences of such a misdiagnosis would have been even more dire, since corticosteroid medication assigned to a person with a vulnerability to cortisol would readily enhance the already manifest psychosis.

Finally, this mechanism provides a good example of an instance where two different drugs administered to patients with similar outward symptoms, may achieve a similar clinical outcome. This is not the only psychiatric context in which such questions have been raised. A more elaborate model could perhaps encompass other such phenomena, such as the similar effects that dopamine agonists and dopamine antagonists seem to have as treatment to certain psychotic symptoms~\cite{Strange}.

\section{Limitations and future work}

\subsection{The limits and advantages of modeling}

One specific problem that has been pointed out in both our previous and current paradigms is that the psychotic symptoms we describe and model are not necessarily specific to schizophrenia, and could be considered common to other conditions of emotional dysregulation of the brain. However, this problem is more of a linguistic tautology, since it is the psychiatric diagnostic itself that we are challenging. 

A more serious drawback comes from the fact that the physiology described by models such as ours can only capture a simplified and approximate picture. At this point in psychiatric research, modeling cannot be considered the final answer, but rather only a step towards a better understanding of the problem and its complexity. Clearly, there can always be a better model.

However, a more elaborate model would need to be supported with more precise physiological information. In this context, the state of the current experimental data is at a crossroads. Take for instance our paradigm of immunity-based psychosis. On one hand, research of both physiology of mental illness and autoimmunity mechanisms has accomplished great progress in the past few decades. On the other hand, knowledge of both autoimmunity and schizophrenia --- specifically that of antibodies against brain structures of neurons and glial cells --- is still controversial. The first steps to understanding how antigen-antibody pathology might play in mental illness has been established by animal research~\cite{Williams}, which is debatably relevant to human pathophysiology. In clinical research, while some found serum antibrain antibodies in 28-95$\%$ of studied schizophrenics~\cite{Kuzn}~\cite{Fessel}~\cite{DeLisi1985}~\cite{Henneberg}, others were unable to reproduce these findings~\cite{Rubin}~\cite{Logan}~\cite{Ehrnst}.

We suggest that this quandary can be solved only as experimental research comes together with the model theoretical and computational efforts, completing and validating each other, to produce a stronger hypothesis of the mechanisms that govern autoimmunity and mental illness. Indeed, psychiatric modeling is a science that should evolve interactively with the progress in experimental methods. New techniques of measuring physiological parameters such as brain activation (fMRI, NIRS, MEG) or autoimmunity~\cite{Rothermundt}~\cite{Ely} offer better suited means of experimental investigation. In some contexts (such as autoimmune dysregulation and its psychiatric effects), these techniques have not yet been exploited at their fullest~\cite{Cohen}. 

Such methods have spurred a new culture of parameter-identification and other techniques of quantitative evaluation from data (such as Dynamic Causal Modeling~\cite{Friston}, or computation of dynamical invariants~\cite{Rosenstein}). These are appropriate to be used and should be used to create and validate more realistic theoretical models .

\subsection{The future of diagnosis}

Based on the novel quantitative approaches to psychophysiology and psychopathology, one concept which has recently started to gain ground is temporal architecture profiling~\cite{Radulescu}~\cite{Peled}. Profiling could complement, perhaps even totally eliminate the current diagnosis assignment, which is not only unreliable and incompatible with multiple etiologies, but also socially undesirable and stigmatizing. This could finally mean assembling a quantitative assessment toolbox for schizophrenia. The dynamical ``brain profile'' of a particular patient or high risk individual could be created --- possibly from a set of clinically measurable parameters --- describing physiological features such as brain interactions, or stress vulnerability. This profile could then be compared against a multidimensional, continuous profile chart, constructed based on common statistics. The individual would thus be placed in the right locus of risk/vulnerability, which would facilitate predictions and help assign the most appropriate treatment.

Such theories have already started to emerge in psychiatric literature. Peled~\cite{Peled}, for example, proposes a brain profiling chart based on three dimensions: neural complexity disorders, neuronal resilience insufficiency and context-sensitive processing decline. Relevant equations would be used to calculate and normalize the different values attributed to relevant brain disturbances. The first dimension relates to disturbances occurring to fast neuronal activations and incorporates connectivity and hierarchical imbalances. The second dimension relates to disturbances that alter long-term synaptic modulations, and incorporates disturbances to optimization within neuronal circuitry. Finally the third dimension refers to the level of internal representations. 

In the same direction of thinking, our previous work suggests an example of rudimentary profiling based on two dimensions: the amygdalar self-inhibition $\mu_1$ and the level of nonlinearity of the prefrontal-limbic system, measured by the Lyapunov number $\sigma$~\cite{Radulescu}. Our current model suggests that a better classification could be obtained from considering a multi-dimensional profile, for example in the parameter space $(\mu_1, a_1, a_2, \xi, \beta)$.

This view of psychiatric diagnosis is still in its cradle, but it offers great promise and testable predictions about the etiology of mental disorders. It studies are time trajectories, which intrinsically encompass the clinical history of the patient. It is also brain-related, so it relies on brain imaging investigations which are much more precise than behavioral assessments.
In short, brain profiling diagnosis could be a bold new step towards a reformed psychiatry.



\begin{thebibliography}{99}

\bibitem{Anan}Ananth H, et. al, 2002. Cortical and subcortical gray matter abnormalities in schizophrenia determined through structural magnetic resonance imaging with optimized volumetric voxel-based morphometry. Am J of
Psych, 159.

\bibitem{Andr}Andreasen N, et. al, 1986. Structural abnormalities in the frontal system in schizophrenia. A magnetic resonance imaging study. Arch of Gen Psychiatry, 43.

\bibitem{page}Arthritis Research Campaign. http://www.arc.org.uk/news/pressreleases/awareness/deadly.asp. Released Jan 1998. Accessed 08/17/08.

\bibitem{Awouters}Awouters FH, Lewi PJ, 2007. Forty years of antipsychotic Drug research--from haloperidol to paliperidone--with Dr. Paul Janssen. Arzneimittelforschung. 57 (10) 625-632.

\bibitem{Bongioanni}Bongioanni P, 1993. The bi- and unilateral correlations between the nervous system and the immune system. Minerva Med. 84 (7-8) 365-381. 

\bibitem{Brady}Brady AM, O'Donnell P, 2004. Dopaminergic modulation of prefrontal cortical input to nucleus accumbens neurons \emph{in vivo}. J Neurosci. 24 (5) 1040-1049.

\bibitem{Bresnihan}Bresnihan B, Oliver M, Williams B, Hughes GR, 1979. An antineuronal antibody cross-reacting with erythrocytes and lymphocytes in systemic lupus erythematosus. Arthritis Rheum. 1979. 22 (4) 313-20.

\bibitem{BT}Bruce I, Turetsky MD, et. al, 2002. Memory-delineated subtypes of schizophrenia: relationship to clinical, neuroanatomical, and
neurophysiological measures. Neuropsychology, 16.

\bibitem{Cerqueira}Cerqueira JJ, Mailliet F, Almeida OFX, Jay TM, Sousa N, 2007. The Prefrontal Cortex as a Key Target of the Maladaptive Response to Stress. J Neurosci. 27 (11) 2781-2787.

\bibitem{Cohen}Cohen W, Roberts WN, Levenson JL. Psychiatric aspects of SLE. In: Lahita R, ed. Systemic Lupus Erythematosis. 4th ed. San Diego, CA: Academic Press; 2004:785-825.

\bibitem{Crow}Crow TJ, Molecular pathology of schizophrenia: more than one disease process? Br Med J. 280 (6207) 66–68. 

\bibitem{DeLisi1985}DeLisi LE, Weber RJ, Pert CB, 1985. Are there antibodies against brain in sera from schizophrenic patient? Review and prospectus. Biol Psychiatry. 20; 110-115.

\bibitem{Denburg}Denburg SD, Carbotte RM, Denburg JA, 1994. Corticosteroids and neuropsychological functioning in patients with systemic lupus erythematosus. Arthritis Rheum. 37 (9) 1311-1320.

\bibitem{DSM}``Diagnostic and Statistical Manual of Mental Disorders - Fourth Edition (DSM-IV)'', American Psychiatric Association, Washington D.C., 1994.

\bibitem{Edwards}Edwards J, Cambridge J, 1998. Rheumatoid arthritis and autoimmunity: a new approach to their cause and how long term cure might be achieved. http://www.ucl.ac.uk/~regfjxe/Arthritis.htm. Accessed 08/17/08.

\bibitem{Ehrnst}Ehrnst A, Wiesel FA, Bjerkenstedt L, Tribukait B, Jonsson J, 1982. Failure to detect immunologic stigmata in schizophrenia. Neuropsychology. 8; 169-171.

\bibitem{Ely}Ely LK, Burrows SR, Purcell AW, Rossjohn J, McCluskey J, 2008. T-cells behaving badly: structural insights into alloreactivity and autoimmunity. Curr Opin Immunol. Epub.

\bibitem{Fessel}Fessel WJ, 1962. Autoimmunity and mental illness. Arch Gen Psychiatry. 6; 320-323.

\bibitem{Grace}Floresco SB, Todd CL, Grace AA, 2001. Glutamatergic afferents from the hippocampus to the nucleus accumbens regulate activity of ventral tegmental area dopamine neurons. J Neurosci. 21 (13) 4915-4922.

\bibitem{Ganguli}Ganguli R, Rabin BS, Kelly RH, Lyte M, Ragu U, 1987. Clinical and laboratory evidence of autoimmunity in acute schizophrenia. Ann N Y Acad Sci. 496; 676-685.

\bibitem{Gaughran}Gaughran F, Welch J in Handbook of Neurochemistry and Molecular Neurobiology Neuroimmunology, by Abel Lajtha, Armen Galoyan and Hugo O. Besedovsky.

\bibitem{Gharavi}Gharavi AE, 2001. Anticardiolipin syndrome: antiphospholipid syndrome. Clin Med. 1; 14-17.

\bibitem{Haidemenos}Haidemenos A, Kontis D, Gazi A, Kallai E, Allin M, Lucia B, 2007. Plasma homocysteine, folate and B12 in chronic schizophrenia. Prog Neuropsychopharmacol Biol Psychiatry. 31 (6) 1289-1296.

\bibitem{Harte}Harte M, O'Connor WT, 2004. Evidence for a differential medial prefrontal dopamine D1 and D2 receptor regulation of local and ventral tegmental glutamate and GABA release: A dual probe microdialysis study in the awake rat. Brain Research. 1017; 120-129.

\bibitem{Henneberg}Henneberg AE, Horter S, Ruffert S, 1994. Increased prevalence of antibrain antibodies in the sera from schizophrenic patients. Schizophr Res. 14 (1) 15-22.

\bibitem{Herbert}Herbert J, Goodyer IM, Grossman AB, Hastings MH, de Kloet ER, Lightman SL, Lupien SJ, Roozendaal B, Seckl JR, 2006. Do corticosteroids damage the brain? J Neuroendocrinol. 18; 393-411. 

\bibitem{Jov}Jovanova-Nesic K, Shoenfeld Y, 2007. Autoimmunity in the brain: the pathogenesis insight from cell biology. Ann N Y Acad Sci. 1107; 142-154. 

\bibitem{Kap}Kapur S, Mamo D. Half a century of antipsychotics and still a central role for dopamine D2 receptors. Prog Neuropsychopharmacol Biol.

\bibitem{Kelly}Kelly RH, Ganguli R, Rabin BS, 1987. Antibody to discrete areas of the brain in normal individuals and patients with schizophrenia. Biol Psychiatry. 22; 1488-1491.

\bibitem{Friston}Kiebel SJ, Kloppel S, Weiskopf N, Friston KJ, 2007. Dynamic causal modeling: A generative model of slice timing in fMRI. NeuroImage 34 (4) 1487-1496.

\bibitem{Kipnis}Kipnis J, Cardon M, Strous RD, Schwartz M, 2006. Loss of autoimmune T cells correlates with brain diseases: possible implications for schizophrenia. Trends Mol Med. 12 (3) 107-112.

\bibitem{Kuzn}Kuznetoza NI, Semenov SF, 1961. Detection of antibrain antibodies in the sera of patients with neuropsychiatric disorders. Zh Nevropatol Psikhiatr. 61; 869-873.

\bibitem{Lau}Lau CG, Zukin RS, 2007. NMDA receptor trafficking in synaptic plasticity and neuropsychiatric disorders. Nat Rev Neurosci. 8 (6) 413-426.

\bibitem{Law}Lawrie SM, et. al, 2003. Structural and functional abnormalities of the amygdala in schizophrenia. Annals of the New York Academy of Sciences, 985.

\bibitem{LeDoux2003}LeDoux J, 2003. The emotional brain, fear, and the amygdala. Cell Mol Neurobiol. 23 (4-5) 727-738.

\bibitem{Lisman}Lisman JE, Grace AA, 2005. The hippocampal-VTA loop: controlling the entry of Information into long-yerm memory. Neuron. 46; 703-713.

\bibitem{Logan}Logan DG, Deodhar SD, 1970. Schizophrenia an immunologic disorder. J Am Med Assoc. 212; 1703-1704.

\bibitem{Medoff}Medoff DR, Holcomb HH, Lahti AC, Tamminga CA, 2001. Probing the human hippocampus using rCBF: contrasts in schizophrenia. Hippocampus. 11 (5) 543-550.

\bibitem{Moskowitz}Moskowitz N, 1989. Systemic lupus erythematosus of the central nervous system: 2. Molecular theories and models for mental disease. Mt Siani J Med.

\bibitem{MPY}Mujica-Parodi LR, Yeregani V, et. al, 2005. Nonlinear complexity
and spectral analyses of heart rate variability in medicated and unmedicated patients with schizophrenia. Neuropsychobiology, 51.

\bibitem{Myin}Myin-Germeys I, Delespaul P, van Os J, 2005. Behavioural sensitization to daily life stress in psychosis. Psychol. Med. 35 (5) 733-741.

\bibitem{Nuech}Nuechterlein KH, et. al, 1994. The vulnerability/stress model of schizophrenic relapse: a longitudinal study. Acta Psychiatrica
Scandinavica. Supplementum, 382.

\bibitem{Omdal}Omdal R, Brokstad K, Waterloo K, Koldingsnes W, Jonsson R, Mellgren SI, 2005. Neuropsychiatric disturbances in SLE are associated with antibodies against NMDA receptors. Eur J Neurol. 12 (5) 392-398.

\bibitem{Ozcan}Ozcan O, Ipcioglu OM, Gultepe M, Basogglu C, 2008. Altered red cell membrane compositions related to functional vitamin B(12) deficiency manifested by elevated urine methylmalonic acid concentrations in patients with schizophrenia. Ann Clin Biochem. 45 (Pt 1) 44-49.

\bibitem{Pavlides}Pavlides C, Nivon L, McEwen B, 2002. Effects of chronic stress on hippocampal long-term potentiation. Hippocampus. 12 (2) 245-257.

\bibitem{Peled}Peled A, 2006. Brain profiling and clinical-neuroscience. Medical Hypotheses. 67 (4) 941-946.

\bibitem{Preston}Preston AR, 2005.  Hippocampal function, declarative memory, and schizophrenia: anatomic and functional neuroimaging considerations. Curr Neurol Neurosci Rep. 5 (4) 249-256.

\bibitem{Radulescu}R\v{a}dulescu A, 2008. Schizophrenia -- a parameters' game? J Theor Biol. 

\bibitem{Rits}Ritsner M, et. al, 2004. Elevation of the cortisol/dehydroepiandrosterone ratio in schizophrenia patients. Eur Neuropsychopharmacol, 14.

\bibitem{Rosenstein}Rosenstein MT, Collins JJ, De Luca CJ, 1993. A practical method for calculating largest Lyapunov exponents from small data sets. Psysica D. 65 (1-2) 117-134.

\bibitem{Rothermundt}Rothermundt M, Arolt V, Bayer TA, 2001. Review of immunological and immunopathological findings in schizophrenia. 

\bibitem{Rubin}Rubin RT, 1965. Investigation of precipitins to human brain in sera of psychotic patients. Br J Psychiatry. 111; 1003-1106.

\bibitem{Sanna}Sanna G, Bertolaccini ML, Khamashta MA, 2008. Neuropsychiatric involvement in systemic lupus erythematosus: current therapeutic approach. Curr Pharm Des.14 (13) 1261-1269.

\bibitem{Sap}Sapolsky RM, Plotsky PM, 1990. Hypercortisolism and its possible neural bases. Biol Psychiatry. 27 (9) 937-952.

\bibitem{Scolding}Scolding NJ, Joseph FG, 2002. The neuropathology and pathogenesis of systemic lupus erithematosus. Neuropathol Appl Neurobiol. 28, 173-189.

\bibitem{Seamans}Seamans JK, Gorelova N, Durstewitz D, Yang CR, 2001. Bidirectional dopamine modulation of GABAergic inhibition in prefrontal cortical pyramidal neurons. Journal of Neuroscience. 21 (10) 3628-3638.

\bibitem{Shanahan}Shanahan JC, Kimberly RP. Corticosteroid use in SLE. In: Lahita R, ed. Systemic Lupus Erythematosis. 4th ed. San Diego, CA: Academic Press; 2004:785-825.

\bibitem{Sotres2004}Sotres-Bayon F, Bush D, LeDoux JE, 2004. Emotional perseveration: an update on prefrontal -- amygdala interactions in fear extinction. Learning Memory. 11, 525-535.

\bibitem{Staa}Staal WG, et. al, 2001. Structural brain abnormalities in chronic schizophrenia at the extremes of the outcome spectrum. Am J of Psychiatry, 15. 

\bibitem{Stamm}Stamm R, Buhler KE, 2001. Concepts of vulnerability of psychiatric diseases. Fortschr Neurol Psychiatr. 69 (7) 300-309.

\bibitem{Stevenson}Stevenson CW, Gratton A, 2003. Basolateral amygdala modulation of the nucleus accumbens dopamine response to stress: role of the medial prefrontal cortex. European Journal of Neuroscience. 17; 1287-1295.

\bibitem{Strange}Strange PG, 2008. Antipsychotic drug action: antagonism, inverse agonism or partial agonism. Trends Pharmacol Sci. 29 (6) 314-321.

\bibitem{Tamm}Tamminga CA, 2006. The neurobiology of cognition in schizophrenia. J Clin Psychiatry. 67 Suppl 9, 9-13; discussion 36-42.

\bibitem{Valesini}Valesini G, Alessandri C, Celestino D, Conti F., 2006. Anti-endothelial antibodies and neuropsychiatric systemic lupus erythematosus. Ann N Y Acad Sci. 1069:118-28. 

\bibitem{Volk}Volk DW, Lewis DA, 2002.  Impaired prefrontal inhibition in schizophrenia: relevance for cognitive dysfunction. Physiology and Behavior. 77 (4) 501-505.

\bibitem{Walker}Walker E, ­Mittal V, Tessner K, 2008. Stress and the hypothalamic pituitary adrenal axis in the developmental course of schizophrenia. Annual Review of Clinical Psychology. 4; 189-216.

\bibitem{Weinberger}Weinberger D, McClure R, 2002. Neurotoxicity, neuroplasticity and Magnetic Resonance Imaging morphometry. Arch Gen Psychiatry. 59, 553-558.

\bibitem{Westerink}Westerink BHC, Enrico P, Feimann J, de Vries JB, 1998. The pharmacology of mesocortical dopamine neurons: a dual-probe microdialysis study in the ventral tegmental area and prefrontal cortex of the rat brain. J Pharmacol Exper Therapeutics. 285 (1) 143-154.

\bibitem{Williams}Williams CA Jr, Schupf N, 1977. Antigen-antibody reactions in rat brain sites induce transient changes in drinking behavior. Science. 196 (4287) 328-330. 

\bibitem{Wong}Wong EYH, Herbert J, 2005. Roles of mineralocorticoid and glucocorticoid receptors in the regulation of progenitor proliferation in the adult hippocampus. European J Neurosci. 22; 785-792.

\bibitem{Yang}Yang ZW, Chengappa NR, Shurin G, Brar JS, Rabin BS, Gubbi AV, Ganguli R, 1994. An association between anti-hippocampal antibody concentration and lymphocyte production of IL-2 in patients with schizophrenia. Psychol Med. 24; 449-455.

\bibitem{Zhang}Zhang Y, Behrens M, Lisman JE, 2008. Prolonged Exposure to NMDAR Antagonist Suppresses Inhibitory Synaptic Transmission in Prefrontal Cortex. J Neurophysiol.

\end{thebibliography}
\end{document}